\documentclass[runningheads]{llncs}
\usepackage{amsmath,amssymb,amsfonts}
\usepackage{algorithmic}
\usepackage{graphicx}
\usepackage{float}
\usepackage{multirow}
\usepackage{url}
\usepackage{adjustbox}
\usepackage{xcolor}
\usepackage{textcomp}
\usepackage[utf8]{inputenc}
\usepackage{longtable}

\begin{document}
\def\SB#1{\textsubscript{#1}}
\title{Self-Supervised MRI Reconstruction with Unrolled Diffusion Models}

\author{Yilmaz Korkmaz\inst{1} \and
Tolga Cukur\inst{2,3} \and
Vishal M. Patel\inst{1}}
% index{Korkmaz, Yilmaz} 
% index{Cukur, Tolga}
% index{Patel, Vishal M.}
% corresponding author: Yilmaz Korkmaz
\authorrunning{Yilmaz Korkmaz et al.}

\institute{Johns Hopkins University, MD, USA \and
Bilkent University, Ankara, Turkey \and National Magnetic Resonance Research Center (UMRAM), Ankara, Turkey
\email{\{ykorkma1,vpatel36\}@jhu.edu, cukur@ee.bilkent.edu.tr}}

\maketitle              % typeset the header of the contribution
\begin{abstract}
Magnetic Resonance Imaging (MRI) produces excellent soft tissue contrast, albeit it is an inherently slow imaging modality. Promising deep learning methods have recently been proposed to reconstruct accelerated MRI scans. However, existing methods still suffer from various limitations regarding image fidelity, contextual sensitivity, and reliance on fully-sampled acquisitions for model training. To comprehensively address these limitations, we propose a novel self-supervised deep reconstruction model, named Self-Supervised Diffusion Reconstruction (SSDiffRecon). SSDiffRecon expresses a conditional diffusion process as an unrolled architecture that interleaves cross-attention transformers for reverse diffusion steps with data-consistency blocks for physics-driven processing. Unlike recent diffusion methods for MRI reconstruction, a self-supervision strategy is adopted to train SSDiffRecon using only undersampled k-space data. Comprehensive experiments on public brain MR datasets demonstrates the superiority of SSDiffRecon against state-of-the-art supervised, and self-supervised baselines in terms of reconstruction speed and quality. 
Implementation will be available at \url{https://github.com/yilmazkorkmaz1/SSDiffRecon}.
\end{abstract}

\section{Introduction}

Magnetic Resonance Imaging (MRI) is one of the most widely used imaging modalities due to its excellent soft tissue contrast, but it has prolonged and costly scan sessions. Therefore, accelerated MRI methods are  needed to improve its clinical utilization. Acceleration through undersampled acquisitions of a subset of k-space samples (i.e., Fourier domain coefficients) results in aliasing artifacts \cite{lustig2007sparse,haldar2010compressed,patel2011gradient}. Many promising deep-learning methods have been proposed to reconstruct images by suppressing aliasing artifacts \cite{Wang2016,Kwon2017,Schlemper2017,lee2018deep,Zhu2018,MoDl,rgan,Yu2018c,Mardani2019b,hammernik2021motion,bakker2022learning,huang2022rethinking,guo2023reference,guo2021over,guo2021multi}. However, many existing methods are limited by suboptimal capture of the data distribution, poor contextual sensitivity, and reliance on fully-sampled acquisitions for model training \cite{KnollGeneralization,Grappa_net,rgan}. 

A recently emergent framework for learning data distributions in computer vision is based on diffusion models \cite{ho2020denoising,nichol2021improved}. Several recent studies have considered diffusion-based MRI reconstructions, where either an unconditional or a conditional diffusion model is trained to generate images and reconstruction is achieved by later injecting data-consistency projections in between diffusion steps during inference \cite{xie2022measurement,cao2022high,cao2022accelerating,peng2022towards,dar2022adaptive}. While promising results have been reported, these diffusion methods can show limited reliability due to omission of physical constraints during training, and undesirable reliance on fully-sampled images. There is a more recent work that tried to mitigate fully-sampled data needs by Cui et al. \cite{cui2022self}. In this work authors proposed a two-staged training strategy where a Bayesian network is used to learn the fully-sampled data distribution to train a score model which is then used for conditional sampling. Our model differs from this approach since we trained it end-to-end without allowing error propagation from distinct training sessions.

To overcome mentioned limitations, we propose a novel self-supervised accelerated MRI reconstruction method, called SSDiffRecon. SSDiffRecon leverages a conditional diffusion model that interleaves linear-complexity cross-attention transformer blocks for denoising with data-consistency projections for fidelity to physical constraints. It further adopts self-supervised learning by prediction of masked-out k-space samples in undersampled acquisitions. SSDiffRecon achieves on par performance with supervised baselines while outperforming self-supervised baselines in terms of inference speed and image fidelity. 

\section{Background}
\subsection{Accelerated MRI Reconstruction}
Acceleration in MRI is achieved via undersampling the acquisitions in the Fourier domain as follows
\begin{equation}
\label{eq:sampling}
F_{p} C I = y,
\end{equation}
where $F_{p}$ is the partial Fourier operator, $C$ denotes coil sensitivity maps, $I$ is the MR image and $y$ is partially acquired k-space data. Reconstruction of fully sampled target MR image $I$ from $y$ is an ill-posed problem since the number of unknowns are higher than the number of equations. Supervised deep learning methods try to solve this ill-posed problem using prior knowledge gathered in the offline training sessions as follows
\begin{equation}
\widehat{I}=\underset{I}{\operatorname{argmin}}\frac{1}{2}\|y-F_{p} C I \|^{2} + \lambda(I),
\end{equation}
where $\widehat{I}$ is the reconstruction, and $\lambda(I)$ is the prior knowledge-guided regularization term. In  supervised reconstruction frameworks, prior knowledge is induced from underlying mapping between under- and fully sampled acquisitions.

\subsection{Denoising Diffusion Models}
In diffusion models \cite{ho2020denoising}, Gaussian noise is progressively mapped on the data via a forward noising process
\begin{equation}
\quad q\left(\mathbf{x}_t \mid \mathbf{x}_{t-1}\right)=\mathcal{N}\left(\mathbf{x}_t ; \sqrt{1-\beta_t} \mathbf{x}_{t-1}, \beta_t \mathbf{I}\right),
\end{equation}
where $\beta_t$ refers to the fixed variance schedule. After a sufficient number of forward diffusion steps $(T)$, $x_{t}$ follows a Gaussian distribution. Then, the backward diffusion process is deployed to gradually denoise $x_T$ to get $x_0$ using a deep neural network as a denoiser as follows
\begin{equation}
p_\theta\left(x_{t-1} \mid x_t\right)=\mathcal{N}\left(x_{t-1} ; \epsilon_\theta\left(x_t, t\right), \sigma_t^2 \mathbf{I}\right),
\end{equation}
where $\sigma_t^2=\tilde{\beta}_t=\frac{1-\bar{\alpha}_{t-1}}{1-\bar{\alpha}_t} \beta_t$ and $\epsilon_\theta$ represents the denoising neural network parametrized during backward diffusion and trained using the following loss \cite{ho2020denoising}
\begin{equation}
L(\theta)=\mathbb{E}_{t, \mathbf{x}_0, \epsilon}\left[\left\|\boldsymbol{\epsilon}-\boldsymbol{\epsilon}_\theta\left(\sqrt{\bar{\alpha}_t} \mathbf{x}_0+\sqrt{1-\bar{\alpha}_t} \boldsymbol{\epsilon}, t\right)\right\|^2\right],
\end{equation}
where $\bar{\alpha}_t=\prod_{m=1}^t \alpha_m$, $\alpha_t=1-\beta_t$ and $\epsilon\sim\mathcal{N}(0,I)$. 

\begin{figure*}[!t]
\includegraphics[width=0.99\textwidth]{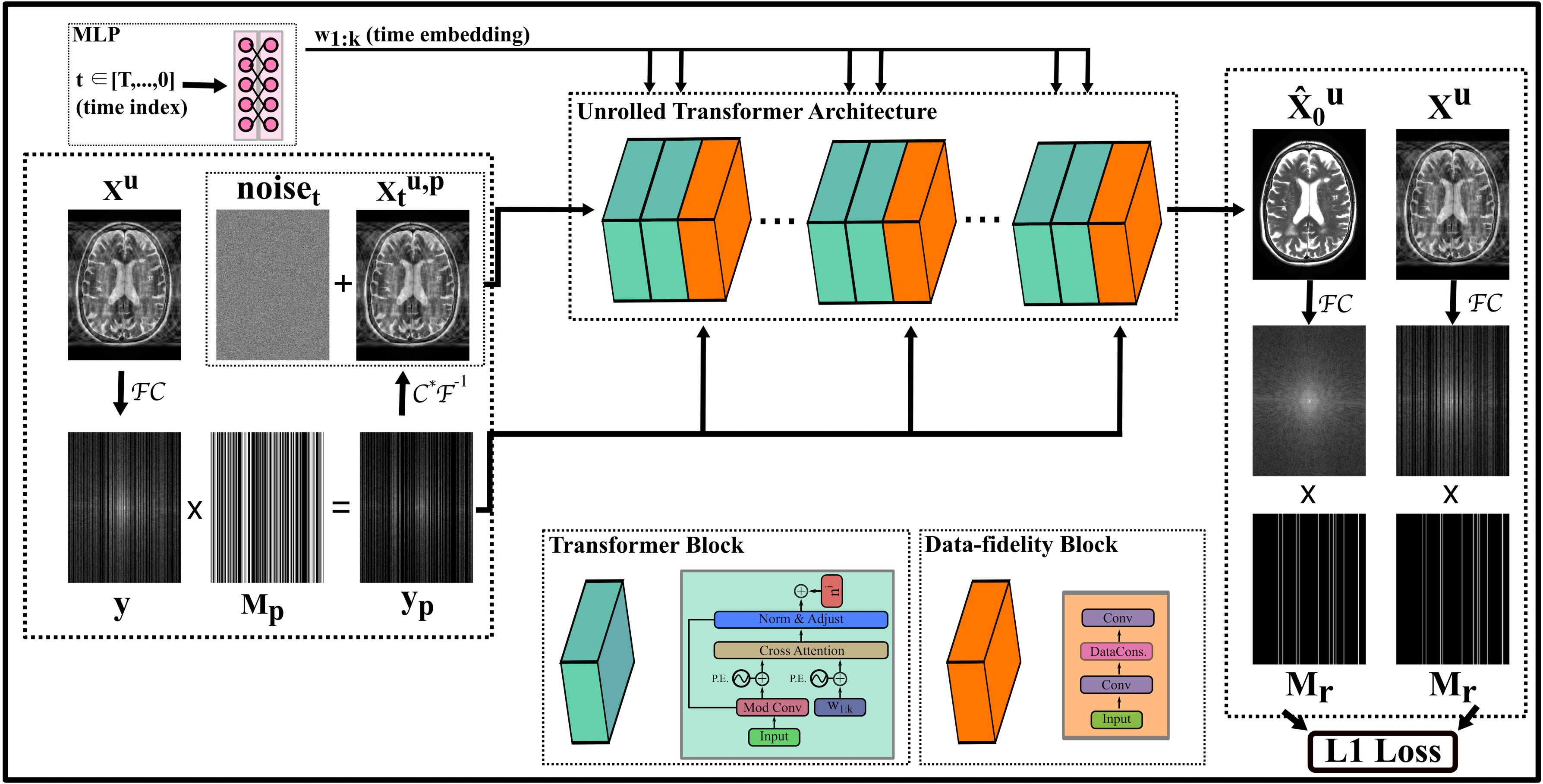}
\caption{Overall training scheme and network architecture. SSDiffRecon utilizes an unrolled physics-guided network as a denoiser in the diffusion process while allowing time index guidance through the Mapper Network via cross-attention transformer layers (shown in green). After two transformer layers, it performs data-consistency (shown in orange). Corresponding noisy input undersampled ($x_t^{u,p}$) and denoised reconstructed images ($\hat{x_0}^{u}$) are shown during training. $L_1$ difference between k-space points in pre-allocated locations ($M_r$) has been utilized as the loss function.}
\label{fig:fig_training}
\end{figure*}

\begin{figure*}[!t]
\includegraphics[width=0.99\textwidth]{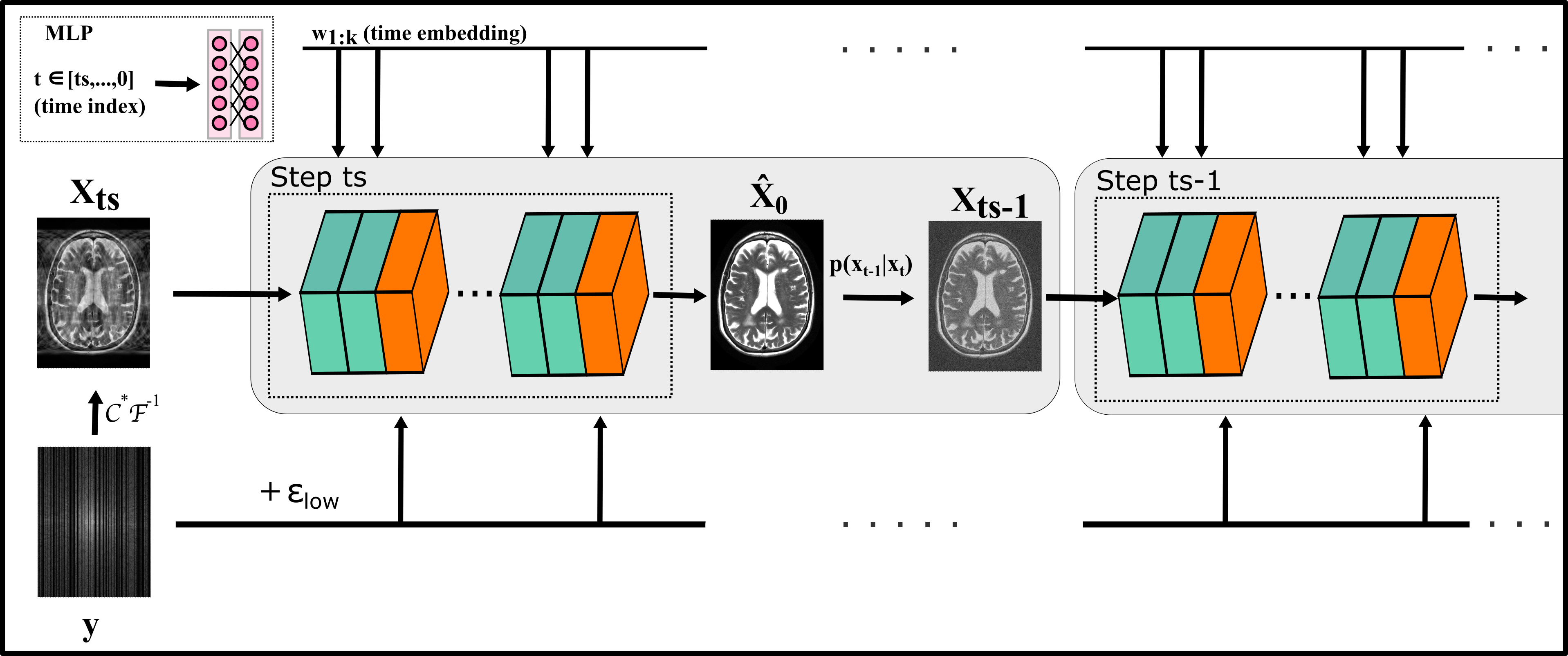}
\caption{Inference scheme. We start from zero-filled reconstruction of undersampled acquisitions and inject a lightweight noise ($\epsilon_{low}$) into them while performing data consistency in the backward diffusion steps to allow a more gradual denoising.}
\label{fig:fig_inference}
\end{figure*}

\section{SSDiffRecon}
 In SSDiffRecon, we utilize a conditional diffusion probabilistic model to reconstruct fully-sampled MR images given undersampled acquisitions as input. The reverse diffusion steps are parametrized using an unrolled transformer architecture as shown in Fig. \ref{fig:fig_training}. To improve adaptation across time steps in the diffusion process, we inject the time-index $t$ via cross-attention transformers as opposed to the original DDPM models that add time embeddings as a bias term. In what follows we describe the training and inference procedures of SSDiffRecon. 
 \paragraph{\textbf{Self-Supervised Training:}} For self-supervised learning, we adopt a k-space masking strategy for diffusion models \cite{yaman2020} as follows
\begin{equation}
    L(\theta) = \|\mathcal{M}_{r}\odot\mathcal{F}(C{x}^{u})-\mathcal{M}_{r}\odot\mathcal{F}(C \hat{x_0}^{u}) \|_1, 
\end{equation}
where $\|\cdot\|_1$ denotes the $L_1$-norm, $\mathcal{F}$ denotes 2D Fourier Transform, $C$ are coil sensitivities, ${x}^{u}$ is the image derived from undersampled acquisitions, and $\mathcal{M}_r$ is the random sub-mask within the main undersampling mask $\mathcal{M}$.  Here $\hat{x_0}^{u}$ is the output of the unrolled denoiser network ($R_{\theta}$) corresponding to the estimate of fully-sampled target image.
\begin{equation}
    \hat{x_0}^{u} = R_{\theta}(x_t^{u,p},y_p,\mathcal{M}_p,C, t), \quad x_t^{u,p} = \sqrt{\bar{\alpha}_t}x^{u,p}+\sqrt{1-\bar{\alpha}_t}\epsilon,
\end{equation}
where $\mathcal{M}_p$ is the sub-mask of the remaining points in $\mathcal{M}$ after excluding $\mathcal{M}_r$, $y_p$ is the further undersampled k-space points using the mask $\mathcal{M}_p$ and $x^{u, p}$ is the zero-filled reconstruction of $y_p$. Training scheme is illustrated in Fig. \ref{fig:fig_training}.

 \paragraph{\textbf{Inference:}} To speed up image sampling, inference starts with zero-filled Fourier reconstruction of the undersampled acquisitions as opposed to a pure noise sample. Conditional diffusion sampling is then performed with the trained diffusion model that iterates through cross-attention transformers for denoising and data-consistency projections. For gradual denoising, we introduce a descending random noise onto the undersampled data within data-consistency layers. Accordingly, the reverse diffusion step at time-index $ts$ is given as
 \begin{equation}
     x_{ts-1} = R_{\theta}(x_{ts}, y_{ts}^{\epsilon}, \mathcal{M},C, ts) + \sigma_{ts} z, \quad y_{ts}^{\epsilon} = \mathcal{F}(\sqrt{\bar{\alpha}_{ts}}x^{u}+\sqrt{1-\bar{\alpha}_{ts}}\epsilon_{low}),
 \end{equation}
where $\epsilon_{low} \sim\mathcal{N}(0,0.1 I)$ and $z\sim\mathcal{N}(0,I)$. Inference procedure is illustrated in Fig. \ref{fig:fig_inference}.

\subsubsection{\textbf{Unrolled Denoising Network $R_{\theta}(.)$}}:
SSDiffRecon deploys an unrolled physics-guided denoiser in the diffusion process instead of UNET as is used in \cite{ho2020denoising}. Our denoiser network consists of the following two fundamental structures as shown in Fig. \ref{fig:fig_training}. The entire network is trained end-to-end.
\begin{enumerate}
    \item Mapper Network
    \item Unrolled Denoising Blocks
\end{enumerate}

\paragraph{\textbf{Mapper network}}: Mapper network is trained to generate local and global latent variables ($w_l$ and $w_g$, respectively) that control the fine and global features in the generated images via cross-attention and instance modulation detailed in later sections. The mapper network is taking time index of the diffusion and extracted label of undersampled image (i.e., undersampling rate and target contrast in multiple contrast dataset) as input and built with 12 fully-connected layers each with 32 neurons.

\paragraph{\textbf{Unrolled Denoising Blocks}}: Each denoising block consists of cross-attention and data-consistency layers sequentially. Let the input of the $j$th denoising block at time instant $t$ be $x_{in, j}^t \in \mathbb{R}^{(h\times w)\times n}$, where $h$ and $w$ denote the height and width of the image, and $n$ denotes the number of feature channels. First, input is modulated with the affine-transformed global latent variable ($w_g \in \mathbb{R}^{32}$) via modulated-convolution adopted from \cite{StyleGAN2}. Assuming that the modulated-convolution kernel is given as $\beta_j$, this operation is expressed as follows
\begin{align}
x_{output,j}^t =\begin{bmatrix}
    \sum_{m} x_{in,j}^{t,m} \circledast \beta_{j}^{m,1} \\
     \vdots \\
     \sum_{m}  x_{in,j}^{t,m} \circledast \beta_{j}^{m,v} \\
\end{bmatrix}, 
\end{align}
where $\beta_{j}^{u,v} \in \mathbb{R}^{3\times 3}$ is the convolution kernel for the $u^{th}$ input channel and the $v^{th}$ output channel, and $m$ is the channel index.
Then, the output of modulated convolution goes into the cross-attention transformer where the attention map $att_{j}^{t}$ is calculated using local latent variables $w_{l}^t$ at time index $t$ as follows
\begin{align}
att_{j}^{t} = softmax\left( \frac{Q_{j}(x_{output, j}^{t}+P.E.)K_j(w_{l}^t+P.E.)^T} {\sqrt{n}}\right) V_j(w_{l}^t),
\end{align} 
where $Q_j(.)$, $K_j(.)$, $V_j(.)$ are queries, keys and values, respectively where each function represents a dense layer with input inside the parenthesis, and $P.E.$ is the positional encoding. 
Then, $x_{output, j}^{t}$ is normalized to zero-mean unit variance and scaled with a learned projection of the attention maps $att_{j}^{t}$ as follows
\begin{align}
x_{output, j}^{t} = \alpha_j(att_{j}) \odot  \left( \frac{x_{output, j}^{t}-\mu(x_{output, j}^{t}) } {\sigma(x_{output, j}^{t})} \right),
\end{align} 
where $\alpha_j(.)$ is the learned scale parameter. After repeating the sequence of cross-attention layer twice, lastly the data-consistency is performed. To perform data-consistency the number of channels in $x_{output, j}^{t}$ is decreased to 2 with an additional convolution layer. Then, 2-channel images are converted, where channels represent real and imaginary components, to complex and data-consistency is applied as follows
\begin{align}
    x_{output, j}^{t}=\mathcal{F}^{-1}\{\mathcal{F}(Cx_{output, j}^{t}) \odot (1-\mathcal{M}_{p}) + \mathcal{F}(Cx^{u}) \odot \mathcal{M}_{p}\},
\end{align}
where $\mathcal{F}^{-1}$ represents the inverse 2D Fourier transform. Then, using another extra convolution, the number of feature maps are increased to $n$ again for the next denoising block.
\paragraph{\textbf{Implementation Details}}: Adam optimizer is used for self-supervised training with $\beta=(0.9, 0.999)$ and learning rate 0.002. Default noise schedule paramaters are taken from \cite{ho2020denoising}. 1000 forward and 5 reverse diffusion steps are used for training and inference respectively with batch size equals to 1. $\mathcal{M}_r$ are sampled from $\mathcal{M}$ using uniform distribution by collecting 5\% of acquired points. We used network snapshots at 445K and 654K steps which corresponds to 28th and 109th epochs for IXI and fastMRI datasets respectively. A single NVIDIA RTX A5000 gpu is used for training and inference.

\section{Experimental Results}
\subsection{Datasets}
Experiments are performed using the following multi-coil and single-coil brain MRI datasets: 
\begin{enumerate}
    \item \textbf{fastMRI}: Reconstruction performance illustrated in multi-coil brain MRI dataset \cite{fastmri}, 100 subjects are used for training, 10 for validation and 40 for testing. Data from multiple sites are included with no common protocol. T\SB{1}-, T\SB{2}- and Flair-weighted acquisitions are considered. GCC \cite{zhang2013coil} is used to decrease the number of coils to 5 to reduce the computational complexity. 
    \item \textbf{IXI}: Reconstruction performance illustrated in single-coil brain MRI data from IXI (http://brain-development.org/ixi-dataset/). T\SB{1}-, T\SB{2}- and PD-weighted acquisitions are considered. In IXI, 25 subjects are used for training, 5 for validation and 10 for testing.
\end{enumerate}
Acquisitions are retrospectively undersampled using variable-density masks. Undersampling masks are generated based on a 2D Gaussian distribution with variance adjusted to obtain acceleration rates of $R = [4, 8]$.

\subsection{Competing Methods}
We compare the performance of SSDiffRecon with the following supervised and self-supervised baselines:
\begin{enumerate}
    \item \textbf{DDPM}: Supervised diffusion-based reconstruction baseline. DDPM is trained with fully sampled MR images and follows a novel k-space sampling approach during inference introduced by Peng et al. \cite{peng2022towards}. 1000 forward and backward diffusion steps are used in training and inference respectively. 
    \item \textbf{self-DDPM}: Self-supervised diffusion-based reconstruction baseline. Self-DDPM is trained using only undersampled MRI acquisitions. Other than training, the inference procedure is identical to the DDPM.
    \item \textbf{D5C5}: Supervised model-based reconstruction baseline. D5C5 is trained using under- and fully sampled paired MR images. Network architecture and training loss are adopted from \cite{qin2018convolutional}.
    \item \textbf{self-D5C5}: Self-supervised model-based reconstruction baseline. Self-D5C5 is trained using the self-supervision approach introduced in \cite{yaman2020} using undersampled acquisitions. The hyperparameters and network architecture are the same as in D5C5.
    \item \textbf{RGAN}: CNN-based reconstruction baseline. RGAN is trained using paired under- and fully sampled MR images. Network architecture and hyperparameters are adapted from \cite{rgan}.
    \item \textbf{self-RGAN}: Self-supervised CNN-based reconstruction baseline. Self-RGAN is trained using the self-supervision loss in \cite{yaman2020} using only undersampled images. Network architecture and other hyperparameters are identical to RGAN.
\end{enumerate}

\subsection{Experiments}
We compared the reconstruction performance using Peak-Signal-to-Noise-Ratio (PSNR, dB) and Structural-Similarity-Index (SSIM, \%) between reconstructions and the ground truth images. Hyperparameter selection for each method is performed via cross-validation to maximize PSNR.

\paragraph{\textbf{Ablation Experiments}}
We perform the following four ablation experiments to show the relative effect of each component in the model on the reconstruction quality as well as  the effect of self-supervision in Table \ref{tab:ablation}.
\begin{enumerate}
    \item {Supervised}: Supervised training of SSDiffRecon using paired under- and fully sampled MR images and pixel-wise loss is performed. Other than training, inference sampling procedures are the same as the SSDiffRecon.
    \item {UNET}: Original UNET architecture in DDPM \cite{ho2020denoising} is trained with the same self-supervised loss as in SSDiffRecon. Other than the denoising network architecture, the training and inference procedures are not changed.
    \item {Without TR}: SSDiffRecon without cross-attention transformer layers is trained and tested. This model only consists of data-consistency and CNN layers. Other than the network, training and inference procedures are not changed.
    \item {Without DC}: SSDiffRecon without the data-consistency layers is trained and tested. This model does not utilize data-consistency but the other training and inference details are the same as the SSDiffRecon.   
\end{enumerate}

% \usepackage{graphicx}
\begin{table}[]
\centering
\caption{Ablation results as avaraged across whole fastMRI test set.}
\resizebox{0.7\textwidth}{!}{%
\begin{tabular}{|c|cc|cc|cc|cc|cc|}
\hline
 &
  \multicolumn{2}{c|}{SSDiffRecon} &
  \multicolumn{2}{c|}{Supervised} &
  \multicolumn{2}{c|}{UNET} &
  \multicolumn{2}{c|}{Without TR} &
  \multicolumn{2}{c|}{Without DC} \\ \hline
 &
  \multicolumn{1}{c|}{PSNR} &
  SSIM &
  \multicolumn{1}{c|}{PSNR} &
  SSIM &
  \multicolumn{1}{c|}{PSNR} &
  SSIM &
  \multicolumn{1}{c|}{PSNR} &
  SSIM &
  \multicolumn{1}{c|}{PSNR} &
  SSIM \\ \hline
fastMRI &
  \multicolumn{1}{c|}{35.9} &
  \textbf{94.1} &
  \multicolumn{1}{c|}{\textbf{36.3}} &
  93.2 &
  \multicolumn{1}{c|}{26.9} &
  84.7 &
  \multicolumn{1}{c|}{35.1} &
  93.8 &
  \multicolumn{1}{c|}{26.4} &
  66.4 \\ \hline
\end{tabular}%
}
\label{tab:ablation}
\end{table}

\begin{table}[]
\caption{Reconstruction performance on the IXI dataset for R = 4 and 8.}
\resizebox{\textwidth}{!}{%
\begin{tabular}{|c|cc|cc|cc|cc|cc|cc|cc|}
\hline
\multicolumn{1}{|l|}{} & \multicolumn{2}{c|}{\textbf{DDPM}} & \multicolumn{2}{c|}{\textbf{D5C5}} & \multicolumn{2}{c|}{\textbf{RGAN}} &  \multicolumn{2}{c|}{\textbf{self-DDPM}} & \multicolumn{2}{c|}{\textbf{Self-D5C5}} & \multicolumn{2}{c|}{\textbf{self-RGAN}} & \multicolumn{2}{c|}{\textbf{SSDiffRecon}}          \\ \hline
\multicolumn{1}{|l|}{} & \multicolumn{1}{c|}{PSNR}  & SSIM  & \multicolumn{1}{c|}{PSNR}  & SSIM  & \multicolumn{1}{c|}{PSNR}  & SSIM  &  \multicolumn{1}{c|}{PSNR}     & SSIM    & \multicolumn{1}{c|}{PSNR}     & SSIM    & \multicolumn{1}{c|}{PSNR}     & SSIM    & \multicolumn{1}{c|}{PSNR}          & SSIM          \\ \hline
T\SB{1}-4x                  & \multicolumn{1}{c|}{39.4}  & 98.8  & \multicolumn{1}{c|}{37.1}  & 97.1  & \multicolumn{1}{c|}{36.8}  & 97.6  &  \multicolumn{1}{c|}{33.6}     & 92.7    & \multicolumn{1}{c|}{39.3}     & 98.4    & \multicolumn{1}{c|}{36.9}     & 97.9    & \multicolumn{1}{c|}{\textbf{42.3}} & \textbf{99.3} \\ \hline
T\SB{1}-8x                  & \multicolumn{1}{c|}{33.0}  & 96.8  & \multicolumn{1}{c|}{30.9}  & 94.0  & \multicolumn{1}{c|}{32.3}  & 95.4  &  \multicolumn{1}{c|}{30.1}     & 90.2    & \multicolumn{1}{c|}{32.7}     & 96.0    & \multicolumn{1}{c|}{31.9}     & 95.9    & \multicolumn{1}{c|}{\textbf{34.6}} & \textbf{97.9} \\ \hline
T\SB{2}-4x                  & \multicolumn{1}{c|}{41.8}  & 98.5  & \multicolumn{1}{c|}{38.9}  & 95.0  & \multicolumn{1}{c|}{38.5}  & 96.6  &  \multicolumn{1}{c|}{35.4}     & 88.1    & \multicolumn{1}{c|}{40.2}     & 97.4    & \multicolumn{1}{c|}{39.0}     & 96.9    & \multicolumn{1}{c|}{\textbf{45.9}} & \textbf{99.1} \\ \hline
T\SB{2}-8x                  & \multicolumn{1}{c|}{36.2}  & 96.2  & \multicolumn{1}{c|}{34.2}  & 91.6  & \multicolumn{1}{c|}{34.6}  & 94.3  &  \multicolumn{1}{c|}{32.9}     & 84.9    & \multicolumn{1}{c|}{34.6}     & 94.5    & \multicolumn{1}{c|}{34.9}     & 94.6    & \multicolumn{1}{c|}{\textbf{39.3}} & \textbf{98.0} \\ \hline
PD-4x               & \multicolumn{1}{c|}{37.1}  & 98.5  & \multicolumn{1}{c|}{36.5}  & 94.7  & \multicolumn{1}{c|}{35.6}  & 96.5  &  \multicolumn{1}{c|}{32.5}     & 88.4    & \multicolumn{1}{c|}{\textbf{39.4}}     & 98.5    & \multicolumn{1}{c|}{36.3}     & 96.8    & \multicolumn{1}{c|}{38.4} & \textbf{99.0} \\ \hline
PD-8x               & \multicolumn{1}{c|}{32.2}  & 96.1  & \multicolumn{1}{c|}{30.1}  & 90.5  & \multicolumn{1}{c|}{31.9}  & 93.9  &  \multicolumn{1}{c|}{29.8}     & 85.0    & \multicolumn{1}{c|}{32.3}     & 94.5    & \multicolumn{1}{c|}{31.8}     & 94.4    & \multicolumn{1}{c|}{\textbf{33.3}} & \textbf{97.8} \\ \hline
\end{tabular}%
}
\label{tab:ixi}
\end{table}

\section{Results}
\begin{figure*}[!t]
\includegraphics[width=0.99\textwidth]{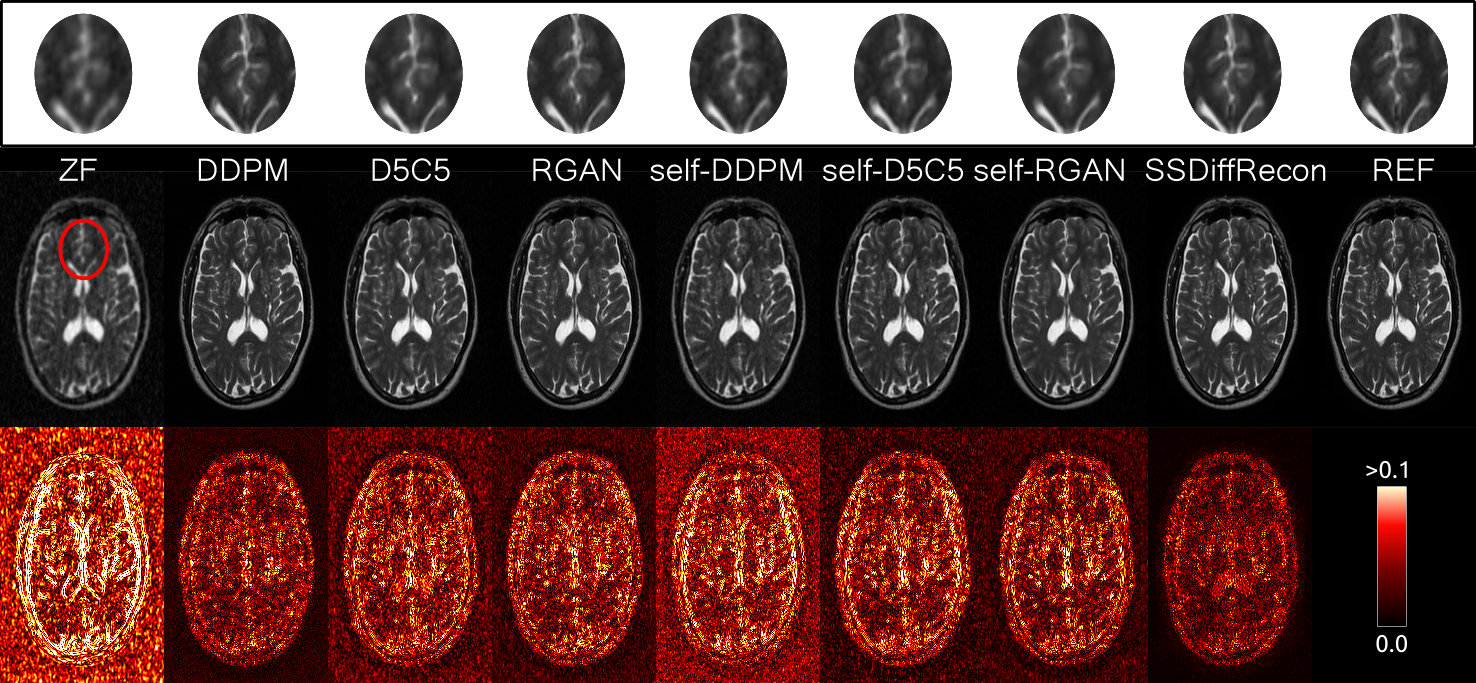}
\caption{Reconstructions of T\SB{2}- weighted images from the IXI dataset, along with the zoomed-in regions on the top and the corresponding error maps underneath.}
\label{fig:fig_2}
\end{figure*}

\begin{figure*}[!t]
\includegraphics[width=0.99\textwidth]{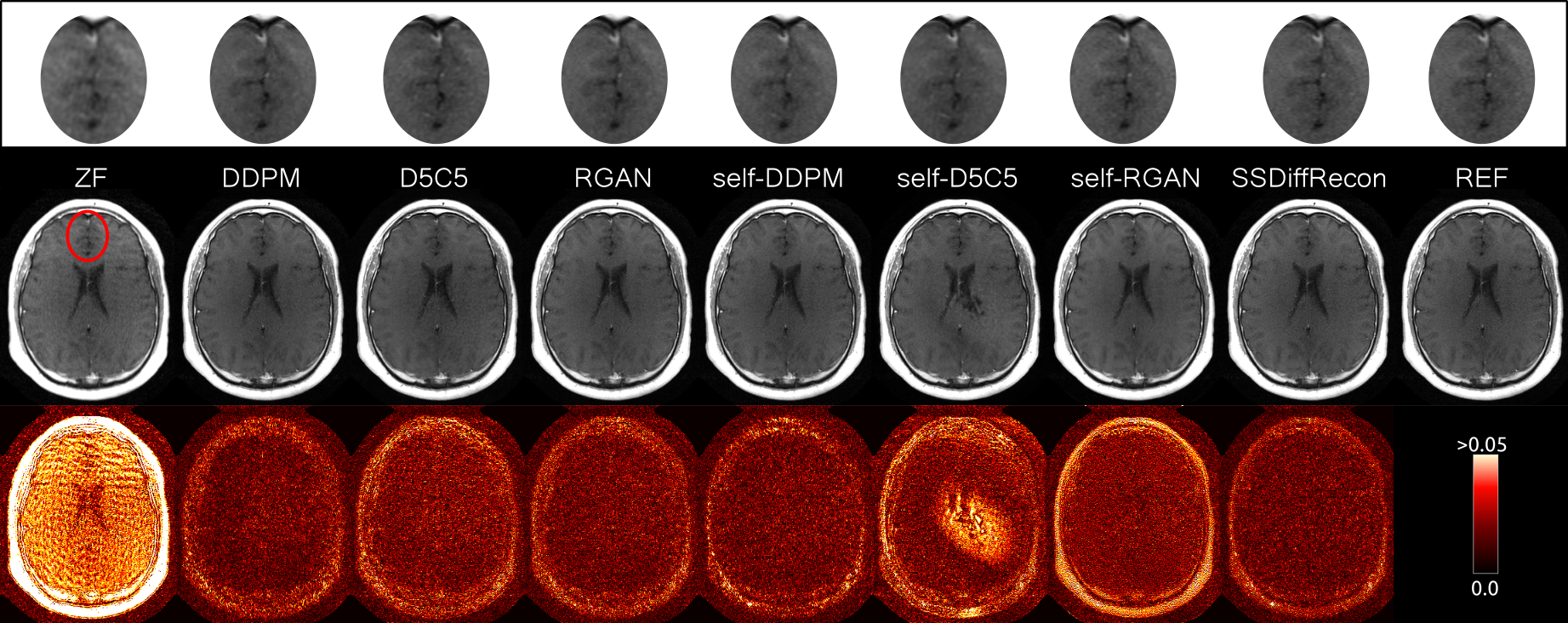}
\caption{Reconstructions of T\SB{1}- weighted images from fastMRI, along with the zoomed-in regions on the top and the corresponding error maps underneath.}
\label{fig:fastmri_fig}
\end{figure*}

The results are presented in two clusters using a single figure and table for each dataset; fastMRI results are presented in Fig. \ref{fig:fastmri_fig} and Table \ref{tab:fastmri}, and IXI results are presented in Fig. \ref{fig:fig_2} and Table \ref{tab:ixi}. The best performed method in each test case is marked in bold in the tables. SSDiffRecon yields 2.55dB more average PSNR and \%1.96 SSIM than the second best self-supervised baseline in IXI, while performing 0.4dB better in terms of PSNR and \%0.25 in terms of SSIM on fastMRI.  Visually, it captured most of the high frequency details while other self-supervised reconstructions suffer from either high noise or blurring artifact. Moreover, visual quality of reconstructions is either very close or better than supervised methods as be seen in the figures. It is also important to note that SSDiffRecon is performing only five backward diffusion steps while regular DDPM perform thousand diffusion steps for an equivalent reconstruction performance. 

\begin{table}[]
\caption{Reconstruction performance on the fastMRI dataset for R = 4 and 8.}
\resizebox{\textwidth}{!}{%
\begin{tabular}{|c|cc|cc|cc|cc|cc|cc|cc|}
\hline
         & \multicolumn{2}{c|}{\textbf{DDPM}}          & \multicolumn{2}{c|}{\textbf{D5C5}} & \multicolumn{2}{c|}{\textbf{RGAN}} &  \multicolumn{2}{c|}{\textbf{self-DDPM}}     & \multicolumn{2}{c|}{\textbf{self-D5C5}} & \multicolumn{2}{c|}{\textbf{Self-RGAN}} & \multicolumn{2}{c|}{\textbf{SSDiffRecon}}            \\ \hline
         & \multicolumn{1}{c|}{PSNR}           & SSIM  & \multicolumn{1}{c|}{PSNR}  & SSIM  & \multicolumn{1}{c|}{PSNR}  & SSIM  &  \multicolumn{1}{c|}{PSNR}  & SSIM           & \multicolumn{1}{c|}{PSNR}     & SSIM    & \multicolumn{1}{c|}{PSNR}     & SSIM    & \multicolumn{1}{c|}{PSNR}           & SSIM           \\ \hline
T\SB{1}-4x    & \multicolumn{1}{c|}{\textbf{40.2}} & 95.3 & \multicolumn{1}{c|}{39.3} & 94.8 & \multicolumn{1}{c|}{39.6} & 95.5 &  \multicolumn{1}{c|}{38.4} & 95.8          & \multicolumn{1}{c|}{38.0}    & 93.1   & \multicolumn{1}{c|}{38.3}    & 95.0   & \multicolumn{1}{c|}{40.1}          & \textbf{96.5} \\ \hline
T\SB{1}-8x    & \multicolumn{1}{c|}{\textbf{36.2}} & 91.7 & \multicolumn{1}{c|}{35.6} & 92.6 & \multicolumn{1}{c|}{36.0} & 92.7 &  \multicolumn{1}{c|}{35.4} & 93.3          & \multicolumn{1}{c|}{34.8}    & 90.5   & \multicolumn{1}{c|}{35.0}    & 92.4   & \multicolumn{1}{c|}{35.1}          & \textbf{93.3} \\ \hline
T\SB{2}-4x    & \multicolumn{1}{c|}{\textbf{38.2}} & 96.0 & \multicolumn{1}{c|}{37.5} & 96.2 & \multicolumn{1}{c|}{36.8} & 95.8 &  \multicolumn{1}{c|}{36.8} & 96.0          & \multicolumn{1}{c|}{37.1}    & 95.6   & \multicolumn{1}{c|}{36.8}    & 95.6   & \multicolumn{1}{c|}{37.7}          & \textbf{96.6} \\ \hline
T\SB{2}-8x    & \multicolumn{1}{c|}{\textbf{34.5}} & 93.0 & \multicolumn{1}{c|}{34.3} & 94.0 & \multicolumn{1}{c|}{33.8} & 93.1 &  \multicolumn{1}{c|}{34.3} & \textbf{94.2} & \multicolumn{1}{c|}{33.8}    & 93.0   & \multicolumn{1}{c|}{34.0}    & 93.4   & \multicolumn{1}{c|}{33.7}          & 93.8          \\ \hline
Flair-4x & \multicolumn{1}{c|}{36.8}          & 93.5 & \multicolumn{1}{c|}{36.2} & 93.3 & \multicolumn{1}{c|}{35.1} & 92.8 &  \multicolumn{1}{c|}{35.7} & 94.1          & \multicolumn{1}{c|}{35.4}    & 91.1   & \multicolumn{1}{c|}{35.3}    & 92.1   & \multicolumn{1}{c|}{\textbf{36.9}} & \textbf{94.6} \\ \hline
Flair-8x & \multicolumn{1}{c|}{\textbf{33.1}} & 87.8 & \multicolumn{1}{c|}{32.7} & 89.1 & \multicolumn{1}{c|}{32.0} & 88.1 &  \multicolumn{1}{c|}{32.5} & 89.6          & \multicolumn{1}{c|}{32.2}    & 86.3   & \multicolumn{1}{c|}{32.1}    & 87.8   & \multicolumn{1}{c|}{32.1}          & \textbf{89.7} \\ \hline
\end{tabular}%
}
\label{tab:fastmri}
\end{table}

\section{Conclusion}
We proposed a novel diffusion-based unrolled architecture for accelerated MRI reconstruction. Our model performs better than self-supervised baselines in a relatively short inference time while performing on-par with the supervised reconstruction methods. Inference time and model complexity analyses are presented in the  supplementary materials.
\subsubsection{Acknowledgement.} This work was supported by NIH R01 grant R01CA276221 and TUBITAK 1001 grant 121E488.

%
% the environments 'definition', 'lemma', 'proposition', 'corollary',
% 'remark', and 'example' are defined in the LLNCS documentclass as well.
%

\clearpage 
\bibliographystyle{splncs04}
\bibliography{main}

\begin{thebibliography}{10}
\providecommand{\url}[1]{\texttt{#1}}
\providecommand{\urlprefix}{URL }
\providecommand{\doi}[1]{https://doi.org/#1}

\bibitem{MoDl}
{Aggarwal}, H.K., {Mani}, M.P., {Jacob}, M.: {MoDL: Model-Based} deep learning architecture for inverse problems. IEEE Transactions on Medical Imaging  \textbf{38}(2),  394--405 (2019)

\bibitem{bakker2022learning}
Bakker, T., Muckley, M., Romero-Soriano, A., Drozdzal, M., Pineda, L.: On learning adaptive acquisition policies for undersampled multi-coil mri reconstruction. arXiv preprint arXiv:2203.16392  (2022)

\bibitem{cao2022high}
Cao, C., Cui, Z.X., Liu, S., Liang, D., Zhu, Y.: High-frequency space diffusion models for accelerated mri. arXiv preprint arXiv:2208.05481  (2022)

\bibitem{cao2022accelerating}
Cao, Y., Wang, L., Zhang, J., Xia, H., Yang, F., Zhu, Y.: Accelerating multi-echo mri in k-space with complex-valued diffusion probabilistic model. In: 2022 16th IEEE International Conference on Signal Processing (ICSP). vol.~1, pp. 479--484. IEEE (2022)

\bibitem{cui2022self}
Cui, Z.X., Cao, C., Liu, S., Zhu, Q., Cheng, J., Wang, H., Zhu, Y., Liang, D.: Self-score: Self-supervised learning on score-based models for mri reconstruction. arXiv preprint arXiv:2209.00835  (2022)

\bibitem{dar2022adaptive}
Dar, S.U., {\"O}zt{\"u}rk, {\c{S}}., Korkmaz, Y., Elmas, G., {\"O}zbey, M., G{\"u}ng{\"o}r, A., {\c{C}}ukur, T.: Adaptive diffusion priors for accelerated mri reconstruction. arXiv preprint arXiv:2207.05876  (2022)

\bibitem{rgan}
Dar, S.U., Yurt, M., Shahdloo, M., Ild{\i}z, M.E., T{\i}naz, B., {\c{C}}ukur, T.: Prior-guided image reconstruction for accelerated multi-contrast {MRI} via generative adversarial networks. IEEE Journal of Selected Topics in Signal Processing  \textbf{14}(6),  1072--1087 (2020)

\bibitem{guo2023reference}
Guo, P., Patel, V.M.: Reference-based mri reconstruction using texture transformer. In: Medical Imaging with Deep Learning (2023)

\bibitem{guo2021over}
Guo, P., Valanarasu, J.M.J., Wang, P., Zhou, J., Jiang, S., Patel, V.M.: Over-and-under complete convolutional rnn for mri reconstruction. In: Medical Image Computing and Computer Assisted Intervention--MICCAI 2021: 24th International Conference, Strasbourg, France, September 27--October 1, 2021, Proceedings, Part VI 24. pp. 13--23. Springer (2021)

\bibitem{guo2021multi}
Guo, P., Wang, P., Zhou, J., Jiang, S., Patel, V.M.: Multi-institutional collaborations for improving deep learning-based magnetic resonance image reconstruction using federated learning. In: Proceedings of the IEEE/CVF Conference on Computer Vision and Pattern Recognition. pp. 2423--2432 (2021)

\bibitem{haldar2010compressed}
Haldar, J.P., Hernando, D., Liang, Z.P.: Compressed-sensing mri with random encoding. IEEE transactions on Medical Imaging  \textbf{30}(4),  893--903 (2010)

\bibitem{hammernik2021motion}
Hammernik, K., Pan, J., Rueckert, D., K{\"u}stner, T.: Motion-guided physics-based learning for cardiac mri reconstruction. In: 2021 55th Asilomar Conference on Signals, Systems, and Computers. pp. 900--907. IEEE (2021)

\bibitem{ho2020denoising}
Ho, J., Jain, A., Abbeel, P.: Denoising diffusion probabilistic models. Advances in Neural Information Processing Systems  \textbf{33},  6840--6851 (2020)

\bibitem{huang2022rethinking}
Huang, W., Li, C., Fan, W., Zhou, Y., Liu, Q., Zheng, H., Wang, S.: Rethinking the optimization process for self-supervised model-driven mri reconstruction. arXiv preprint arXiv:2203.09724  (2022)

\bibitem{StyleGAN2}
Karras, T., Laine, S., Aittala, M., Hellsten, J., Lehtinen, J., Aila, T.: Analyzing and improving the image quality of {StyleGAN}. In: Proceedings of the IEEE/CVF Conference on Computer Vision and Pattern Recognition (CVPR). pp. 8107--8116 (2020)

\bibitem{KnollGeneralization}
Knoll, F., Hammernik, K., Kobler, E., Pock, T., Recht, M.P., Sodickson, D.K.: Assessment of the generalization of learned image reconstruction and the potential for transfer learning. Magnetic Resonance in Medicine  \textbf{81}(1),  116--128 (2019)

\bibitem{fastmri}
Knoll, F., Zbontar, J., Sriram, A., Muckley, M.J., Bruno, M., Defazio, A., Parente, M., Geras, K.J., Katsnelson, J., Chandarana, H., Zhang, Z., Drozdzalv, M., Romero, A., Rabbat, M., Vincent, P., Pinkerton, J., Wang, D., Yakubova, N., Owens, E., Zitnick, C.L., Recht, M.P., Sodickson, D.K., Lui, Y.W.: {fastMRI: A} publicly available raw k-space and {DICOM} dataset of knee images for accelerated {MR} image reconstruction using machine learning. Radiology: Artificial Intelligence  \textbf{2}(1),  e190007 (2020)

\bibitem{Kwon2017}
Kwon, K., Kim, D., Park, H.: {A parallel MR imaging method using multilayer perceptron}. Medical Physics  \textbf{44}(12),  6209--6224 (2017). \doi{10.1002/mp.12600}

\bibitem{lee2018deep}
Lee, D., Yoo, J., Tak, S., Ye, J.C.: Deep residual learning for accelerated mri using magnitude and phase networks. IEEE Transactions on Biomedical Engineering  \textbf{65}(9),  1985--1995 (2018)

\bibitem{lustig2007sparse}
Lustig, M., Donoho, D., Pauly, J.M.: Sparse mri: The application of compressed sensing for rapid mr imaging. Magnetic Resonance in Medicine: An Official Journal of the International Society for Magnetic Resonance in Medicine  \textbf{58}(6),  1182--1195 (2007)

\bibitem{Mardani2019b}
Mardani, M., Gong, E., Cheng, J.Y., Vasanawala, S., Zaharchuk, G., Xing, L., Pauly, J.M.: {Deep generative adversarial neural networks for compressive sensing MRI}. IEEE Transactions on Medical Imaging  \textbf{38}(1),  167--179 (2019)

\bibitem{nichol2021improved}
Nichol, A.Q., Dhariwal, P.: Improved denoising diffusion probabilistic models. In: International Conference on Machine Learning. pp. 8162--8171. PMLR (2021)

\bibitem{patel2011gradient}
Patel, V.M., Maleh, R., Gilbert, A.C., Chellappa, R.: Gradient-based image recovery methods from incomplete fourier measurements. IEEE Transactions on Image Processing  \textbf{21}(1),  94--105 (2011)

\bibitem{peng2022towards}
Peng, C., Guo, P., Zhou, S.K., Patel, V.M., Chellappa, R.: Towards performant and reliable undersampled mr reconstruction via diffusion model sampling. In: Medical Image Computing and Computer Assisted Intervention--MICCAI 2022: 25th International Conference, Singapore, September 18--22, 2022, Proceedings, Part VI. pp. 623--633. Springer (2022)

\bibitem{qin2018convolutional}
Qin, C., Schlemper, J., Caballero, J., Price, A.N., Hajnal, J.V., Rueckert, D.: Convolutional recurrent neural networks for dynamic mr image reconstruction. IEEE transactions on medical imaging  \textbf{38}(1),  280--290 (2018)

\bibitem{Schlemper2017}
Schlemper, J., Caballero, J., Hajnal, J.V., Price, A., Rueckert, D.: {A Deep Cascade of Convolutional Neural Networks for {MR} Image Reconstruction}. In: International Conference on Information Processing in Medical Imaging. pp. 647--658 (2017)

\bibitem{Grappa_net}
Sriram, A., Zbontar, J., Murrell, T., Zitnick, C.L., Defazio, A., Sodickson, D.K.: {GrappaNet}: {Combining} parallel imaging with deep learning for multi-coil {MRI} reconstruction. In: Proceedings of the IEEE/CVF Conference on Computer Vision and Pattern Recognition (CVPR). pp. 14303--14310 (June 2020)

\bibitem{Wang2016}
Wang, S., Su, Z., Ying, L., Peng, X., Zhu, S., Liang, F., Feng, D., Liang, D.: {Accelerating magnetic resonance imaging via deep learning}. In: IEEE 13th International Symposium on Biomedical Imaging (ISBI). pp. 514--517 (2016). \doi{10.1109/ISBI.2016.7493320}

\bibitem{xie2022measurement}
Xie, Y., Li, Q.: Measurement-conditioned denoising diffusion probabilistic model for under-sampled medical image reconstruction. In: Medical Image Computing and Computer Assisted Intervention--MICCAI 2022: 25th International Conference, Singapore, September 18--22, 2022, Proceedings, Part VI. pp. 655--664. Springer (2022)

\bibitem{yaman2020}
Yaman, B., Hosseini, S.A.H., Moeller, S., Ellermann, J., U{\u{g}}urbil, K., Ak{\c{c}}akaya, M.: Self-supervised learning of physics-guided reconstruction neural networks without fully sampled reference data. Magnetic resonance in medicine  \textbf{84}(6),  3172--3191 (2020)

\bibitem{Yu2018c}
Yu, S., Dong, H., Yang, G., Slabaugh, G., Dragotti, P.L., Ye, X., Liu, F., Arridge, S., Keegan, J., Firmin, D., Guo, Y.: {DAGAN: Deep de-aliasing generative adversarial networks for fast compressed sensing MRI reconstruction}. IEEE Transactions on Medical Imaging  \textbf{37}(6),  1310--1321 (2018)

\bibitem{zhang2013coil}
Zhang, T., Pauly, J.M., Vasanawala, S.S., Lustig, M.: Coil compression for accelerated imaging with cartesian sampling. Magnetic resonance in medicine  \textbf{69}(2),  571--582 (2013)

\bibitem{Zhu2018}
Zhu, B., Liu, J.Z., Rosen, B.R., Rosen, M.S.: {Image reconstruction by domain transform manifold learning}. Nature  \textbf{555}(7697),  487--492 (2018)

\end{thebibliography}
\newpage
\section*{Supplementary Materials}

\begin{table}[]
\centering
\caption{Number of trainable parameters are shown for each competing method. Self-supervised baselines (self-DDPM, self-D5C5 and self-RGAN) have the same parameters with the supervised versions since they are using the same network structures. Therefore, only a single value included for each competing network type. As be seen from the numbers, DDPM is the most complex model among all competing methods due to its large denoising UNET.}
\resizebox{.8\textwidth}{!}{%
\begin{tabular}{|c|cl|cl|cl|cl|}
\hline
 &
  \multicolumn{2}{c|}{\textbf{DDPM}} &
  \multicolumn{2}{c|}{\textbf{D5C5}} &
  \multicolumn{2}{c|}{\textbf{RGAN}} &
  \multicolumn{2}{c|}{\textbf{SSDiffRecon}} \\ \hline
\#Trainable Parameters &
  \multicolumn{2}{c|}{164,303,618} &
  \multicolumn{2}{c|}{297,794} &
  \multicolumn{2}{c|}{11,377,154} &
  \multicolumn{2}{c|}{3,292,570} \\ \hline
\end{tabular}%
}

\label{tab:ablation1}
\end{table}
\begin{table}[]
\centering
\caption{Inference times needed to reconstruct a single slice have been presented in seconds. A single NVIDIA RTX A5000 gpu is used in the demonstration. Self-supervised baselines (self-DDPM, self-D5C5 and self-RGAN) have the same inference time with the supervised versions since they are using the same network structures. Therefore, only a single value is included for each network type. RGAN and D5C5 are feed-forward networks, so the lowest inference time is expected in those methods with a relatively poor reconstruction performance. On the other hand, DDPM is utilizing a thousand backward diffusion steps, thus suffering from prolonged inference time.}
\resizebox{.6\textwidth}{!}{%
\begin{tabular}{|c|cl|cl|cl|cl|}
\hline
 &
  \multicolumn{2}{c|}{\textbf{DDPM}} &
  \multicolumn{2}{c|}{\textbf{D5C5}} &
  \multicolumn{2}{c|}{\textbf{RGAN}} &
  \multicolumn{2}{c|}{\textbf{SSDiffRecon}} \\ \hline
Inference Time &
  \multicolumn{2}{c|}{52.270} &
  \multicolumn{2}{c|}{0.045} &
  \multicolumn{2}{c|}{0.004} &
  \multicolumn{2}{c|}{0.219} \\ \hline
\end{tabular}%
}

\label{tab:ablation2}
\end{table}
\null
\vfill
\end{document}